# The Goldilocks zone of governing technology: Leveraging uncertainty for responsible quantum practices

**Authors:**

Miriam Meckel

Institute for Media and Communications Management, University of St. Gallen; St. Gallen, Switzerland, miriam.meckel@unisg.ch

Philipp Hacker

European New School of Digital Studies, European University Viadrina; Frankfurt, Germany, Hacker@europa-uni.de

Léa Steinacker

The Nordic Centre for Internet and Society, Department of Communication and Culture, Norwegian Business School BI; Oslo, Norway, lea.steinacker@gmail.com

Aurelija Lukoseviciene

Faculty of Law, Lund University; Lund, Sweden, aurelija.lukoseviciene@jur.lu.se

Surjo R. Soekadar

Clinical Neurotechnology Laboratory, Department of Psychiatry and Neurosciences, Charité Campus Mitte (CCM), Charité - Universitätsmedizin Berlin; Berlin, Germany, surjo.soekadar@charite.de

Jacob Slosser

Sapien Institute, Center for European, Comparative, and Constitutional Legal Studies, Faculty of Law, University of Copenhagen; Copenhagen, Denmark, connect@sapien.institute

Gina-Maria Pöhlmann

Institute for Media and Communications Management, University of St. Gallen; St. Gallen, Switzerland, Poehlmann, gina-maria.poehlmann@unisg.ch

**Abstract:** Emerging technologies challenge conventional governance approaches, especially when uncertainty is not a temporary obstacle but a foundational feature as in quantum computing. This paper reframes uncertainty from a governance liability to a generative force, using the paradigms of quantum mechanics to propose adaptive, probabilistic frameworks for responsible innovation. We identify three interdependent layers of uncertainty—physical, technical, and societal—central to the evolution of quantum technologies. The proposed Quantum Risk Simulator (QRS) serves as a conceptual example, an imaginative blueprint rather than a prescriptive tool, meant to illustrate how probabilistic reasoning could guide dynamic, uncertainty-based governance. By foregrounding epistemic and ontological ambiguity, and drawing analogies from cognitive neuroscience and predictive processing, we suggest a new model of governance aligned with the probabilistic essence of quantum systems. This model, we argue, is especially promising for the European Union as a third way between laissez-faire innovation and state-led control, offering a flexible yet responsible pathway for regulating quantum and other frontier technologies.

# The Goldilocks zone of governing technology: Leveraging uncertainty for responsible quantum practices

## Introduction

*Collingridge dilemma of technological innovation*

In the early stages of any emerging technology, there is a bargain with uncertainty known as the Collingridge Dilemma (Collingridge, 1980): While the technology is not yet sufficiently developed and widespread, its technological, economic, and wider social impacts remain uncertain. Once firmly established, however, technology becomes difficult to control and shape. Current governance frameworks, like the EU AI Act,[1] reflect this conundrum as they are either shrouded in obscurity or grounded in deterministic categories that presume we can define and contain risk in advance. Classical computing, with its binary determinism and linear logic, has shaped much of our legal and regulatory infrastructure. Regulatory paradigms such as the EU AI Act often mirror these structures, favoring predefined or at least calculable risk categories and rule-based compliance mechanisms. This legacy reinforces the assumption that uncertainty can and should be minimized, rather than constructively engaged. As a contrast to that, one emerging field whose very core rests upon uncertainty to an even greater extent than many others is quantum technology. By nature, quantum technology builds on paradigms where uncertainty is not just inevitable but fundamental (Heisenberg, 1927). We argue that its underpinnings provide unique intellectual insights about the ubiquity of uncertainty in technology governance and the necessity to address it in new ways.

We propose an essential reframing from deterministic to probabilistic technological management for creating *responsible* quantum computing frameworks (Krarup and Horst, 2023), positioning uncertainty as a generative force rather than a barrier. We suggest a proactive engagement with the types of uncertainty inherent in quantum mechanics to inspire adaptive mechanisms for responsible technological evolution. As one example, we illustrate a potential instrument of a dynamic governance approach that integrates probabilistic reasoning. Europe could particularly leverage this approach as a third way of flexible regulation that is based on a dynamic approach compared to the free market momentum (USA) and state dirigisme (China) (Krarup and Horst, 2023; Hine and Floridi, 2024).

*Epistemic and ontological uncertainty*

Human beings, as a species, seek to avoid uncertainty (Slovic, 1987; Smithson, 1989) and have developed sophisticated cognitive abilities to reduce it (Tversky et al., 1982; Johnson et al., 2013; Volz and Gigerenzer, 2012; Sloman, 1996). However, uncertainty is inherent in the human experience and a prerequisite for our evolution. While human uncertainty is grounded in a state of unavailable or unknown information (Walker et al., 2013) in quantum mechanics it is a general property of quantum systems. Uncertainty in quantum computing is not simply the absence of knowledge: paradoxically, in quantum systems, uncertainty can prevail even in situations where ample information is available (Harrigan and Spekkens, 2010). Moreover, while human uncertainty can only gradually be reduced by information intake, quantum states can be controlled in well-designed systems until the act of measurement lets them collapse (Frigg et al., 2011). When we measure a quantum system that exists in a superposition of states, we do not simply reveal a pre-existing property, the act of measurement itself plays a crucial role in determining the outcome. When we perform a measurement, the superposition collapses instantaneously into one of the possible eigenstates of the measured observable (Ding and Chong, 2022). Moreover, the particular outcome we observe is also inherently probabilistic - we cannot, a priori, predict with certainty which state will be realized.

Predictive processing, the neuroscientific concept that interprets the mind as a probabilistic prediction engine, offers a compelling framework for understanding both how human cognition and quantum systems deal with uncertainty (Nave et al., 1994). Just as our brains constantly generate predictions based on limited information to navigate an uncertain world, quantum computing operates at the edge of chaos and uncertainty. In cognitive science, this edge is recognized as fertile ground for creativity and problem solving. As we delve deeper into the quantum realm, we must craft governance approaches that not only acknowledge this fundamental uncertainty but also harness it as a catalyst for innovation and responsible progress. For just as quantum measurement fundamentally changes the state being observed, technological governance interventions inherently transform the technological landscape they seek to regulate.

## Types of uncertainty

Quantum computing entails at least three distinct types of uncertainty: underlying physics, technical quantum superiority, and societal effects. Part of such uncertainty is inherent in these technologies and is exploited to achieve unprecedented advancements; another part, however, represents serious risks to various spheres of human lives. At the same time, these uncertainties can serve as steppingstones to give insight into responsible governance approaches not only for quantum but also other emerging technologies.

*Uncertainty of underlying physics*

Humans struggle to understand the intricacies of the inner workings of many new technologies, but for quantum computing, this lack of understanding expands to a second layer of inexplicability: the physics which seems to defy or even contradict our intuitive logic (Grinbaum, 1994). For instance, the inability to predict with certainty the outcome of a single measurement of a particle in superposition state (Shankar, 1994; Preskill, 2018) challenges our classical notions of determinism and observability (Pusey et al., 2012). Quantum entanglement which describes correlations between quantum particles that persist even when the particles are separated by large distances contradicts our intuitions about locality and causality, and creates interpretative challenges for human observers (Pan et al., 2000).

While such phenomena defy classical physics, they may appear less alien when described metaphorically in human or informational terms. People routinely navigate emotional ambivalence, conflicting intentions, or socially entangled situations. However, it is crucial to distinguish between metaphorical parallels and the formal, physical properties of quantum systems. The unpredictability in human decision-making stems from psychological or contextual complexity and is typically epistemic, reflecting limited knowledge or high-dimensional dynamics. In contrast, in quantum mechanics, uncertainty is not epistemic alone but ontological. It is embedded in reality itself.

The need for precise control of qubits clashes with their inherently probabilistic behavior, creating challenges in achieving reliable and scalable quantum computations. At the same time, the uncertainty we associate with quantum phenomena also arises from the limitations of our classical intuitions in grasping these concepts, as review studies of existing narratives of the technology have shown (Suter et al., 2024; Lukoseviciene, 2025). As we continue to develop quantum technologies, we must navigate both the inherent quantum behaviors and our human perspectives on them.

*Uncertainty of technical quantum superiority*

Another layer of uncertainty is introduced by the nebulous timeline, extent, and applicability of the technical development of quantum computing which some experts expect to, at some point, lead to quantum superiority (Aaronson, 2008; Preskill, 2018). For example, Shor's algorithm efficiently factors large numbers, essential for cracking state-of-the-art asymmetric cryptographic systems (Shor, 1994; Ekert and Josza, 1996) but several million of qubits or even more are estimated to be necessary to break the strongest public key encryption systems currently in use and that is far from what is technologically possible right now (Ichikawa et al., 2024; Scholten et al., 2024).

Notably, the achievement of practical quantum computing superiority in prime factorization threatens to render current asymmetric encryption methods (e.g., the RSA algorithm) obsolete (Ali, 2020), and proactive steps have already been taken to mitigate those risks (Liman and Weber, 2023). Conversely, quantum computing also offers a solution through quantum key distribution (QKD) by vastly enhancing the security of symmetric encryption. This epitomizes the dual nature of quantum advancements as well as the ongoing uncertainty and debate over claims of technical progress and quantum superiority.

Moreover, high error rates continue to complicate the achievement of consistent quantum superiority. Yet recent breakthroughs such as improved fault-tolerant algorithms and hardware advances by companies like Xanadu suggest meaningful progress (Larsen et al., 2025). As outlined in Brun's comprehensive work on quantum error correction (2019), techniques to stabilize qubits are rapidly evolving (see also Lanka et al., 2025). Hybrid quantum-classical algorithms are being developed to leverage the strengths of both quantum and classical computing (Tang and Martonosi, 2024). Despite these promising advancements, the journey from theoretical promise to practical application remains challenging (Campbell, 2024).

*Uncertainty of societal effects*

Compounding this technological complexity, the societal repercussions and consequences of quantum computing encompass privacy, security, economic, and social dimensions to an extent that may supersede previous technologies (de Wolf, 2017). Most obviously, the ability of quantum computers to break current encryption methods poses a significant threat to privacy, as it could expose sensitive data (van Daleen, 2024; DeRose, 2023). Moreover, quantum computing development is capital-intensive, potentially leading to redistribution of capital and knowledge, further concentrating wealth among the few entities capable of investing in this technology (a winner-takes-all market) (Kop et al., 2024; Hoofnagle and Garfinkel, 2022). This extends the social implications of quantum computing to the reinforcement of social inequality – a "quantum divide" (Coenen et al., 2022; Seskir et al., 2023; Gercek and Seskir, 2025). Many other uncertain but possibly detrimental social effects can be anticipated using the logic of quantum technologies as "dual use" in nature (Bostrom and Ćirković, 2008).

*From calculable risk to uncertainty: regulatory consequences*

These three uncertainties generally distinguish quantum computing from classical computing, however, the societal effects of other emerging technologies such as advanced AI based on classical computing are also quite uncertain (Anderlfjung et al., 2023; Hacker et al., 2025). Recent regulatory advances, such as the EU AI Act and the GDPR, are primarily based on the classical paradigm of calculable risk in what is called the risk-based approach (Gellert, 2020; Ebers, 2024). The AI Act, for example, contains various rules on risk assessment and management for AI providers, the implicit assumption being that those risks are indeed quantifiable ex ante, even along the value chain, and that mitigation procedures can be calibrated accordingly (cf. Hacker & Holweg, 2025). Especially for quantum computing and quantum phenomena, such an approach seems incomplete – as it sidelines the more radical uncertainties at play.

## Quantum Risk Simulator

Leveraging a new perspective on uncertainty means navigating the realm between the knowns and unknowns, the borderland between order and chaos that is known as the "Goldilocks Zone".[2] It means combining the right level of certainty to survive with the flexibility of uncertainty, allowing openness to new ideas and perspectives. Such an approach should not introduce a variation of Laplace's demon (Laplace 1902) by attempting to provide a false sense of predictability but rather offer a sophisticated framework for decision-making in the face of the many uncertainties of quantum – and other emerging technologies. Moreover, the process of building, interacting with, and adapting risk mitigation tools should in itself be a governance practice directing the responsible development and deployment of technology.

As a conceptual example of this balance, imagine a governance tool designed to model risk not as a fixed score, but as a shifting probability landscape: the Quantum Risk Simulator (QRS). Such an interdisciplinary tool for dynamically projecting risk probabilities could help policymakers and technologists simulate the severity of potential impacts, assess confidence in the underlying data, and explore regulatory pathways in real time (Woerner and Egger, 2019). It could facilitate a shift to embracing the inherent uncertainties and offer a significant improvement over current AI and technology regulation methods, which often rely on binary, or at least discrete, decision-making in risk assessment.[3,4,5]

By accounting for a full range of degrees of uncertainty a QRS would help stakeholders anticipate and mitigate unforeseen consequences, fostering a proactive rather than reactive approach to technology management and governance. A QRS should reflect the dynamics of quantum systems and the specific technical, societal, and ethical risks they pose. In practice, such a QRS would have to be designed as a comprehensive software framework guided by these principles:

a. *Probabilistic foundation*: Instead of providing definitive risk scores, a QRS should use as inputs and outputs probability distributions of potential risks as appropriate to create dynamic trajectories.
b. *Dynamic updating*: The QRS should be designed to continuously update its models based on new experimental data and theoretical advancements. This ensures the tool remains relevant as our understanding of quantum systems evolves.
c. *Uncertainty quantification*: The simulator should explicitly quantify different types of uncertainties, e.g. aleatoric uncertainty which describes the inherent randomness in quantum systems; epistemic uncertainty, which encompasses the limitations in our current knowledge of quantum phenomena; or model uncertainty as the potential inaccuracies in the simulation model itself.

The QRS would be built as a multi-layered software system, including a risk assessment module integrating probabilistic models, a user interface layer for input and visualization, and an API for integration with other systems and data sources. We suggest a modular and cloud-based design to allow easy updates and expansions, scalable computing resources, collaborative access for researchers and decision-makers, and real-time updates and improvements. It should integrate machine learning techniques to improve risk predictions over time, identify patterns in quantum system behavior, and adapt to new quantum technologies as they emerge. We propose it as a blueprint for analyzing other emerging technologies, from AI to geoengineering.

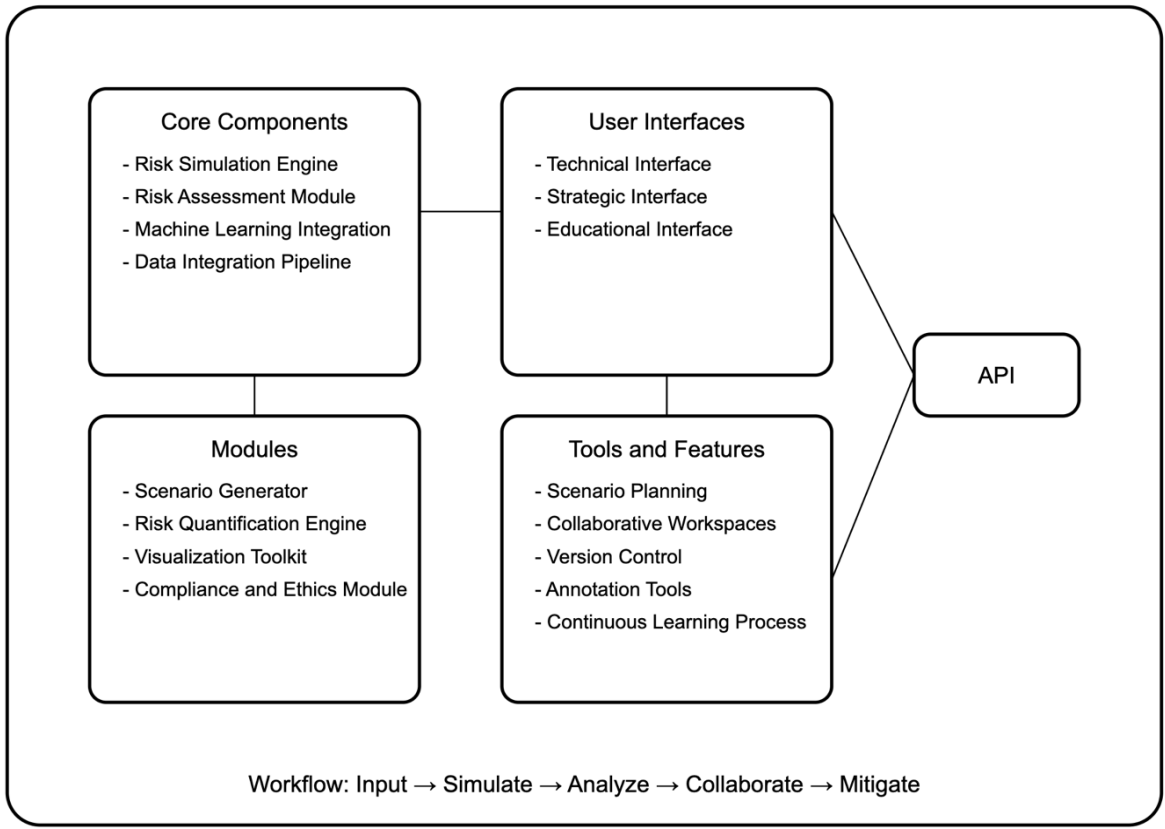

Image 1: Quantum Risk Simulator Cloud Platform

The QRS should provide a multiple user interface with API options for an interdisciplinary group of stakeholders, including engineers, researchers, political strategists, and managers. It would need a controlled and robust data integration pipeline to continuously aggregate new information from research papers, quantum experiments, industry reports, regulatory frameworks and other important documents and data and advanced visualization tools that can represent multi-dimensional probability distributions in an intuitive manner (e.g. heat maps etc.). It should ideally be realized through a public-private partnership, leveraging the innovative capacity of industry alongside the regulatory oversight and public interest considerations of governmental bodies.

Earlier models, such as the World3 system developed for the Club of Rome's *Limits to Growth*, similarly sought to simulate long-term dynamics under uncertainty (Meadows et al, 1972; Herrington, 2021). Unlike these top-down, scenario-based models, the QRS embraces uncertainty as an evolving feature, not a solvable variable. We recognize the inherent limits of modeling complex systems as no simulation can fully predict outcomes at the edge of chaos. Rather, the QRS aims to support reflective governance by continuously updating and exposing assumptions, not resolving them.

Of course, over-reliance on any single risk assessment tool, such as the QRS, could foster a false sense of security in anticipating unforeseen consequences. To avoid complacency, it is crucial to establish periodic evaluations of the QRS and its outcomes. An oversight board, drawing its members from diverse stakeholder groups including scientists, ethicists, policymakers, industry representatives, and civil society organizations, would ensure a balanced and comprehensive approach to risk assessment and mitigation.

**Conclusion**

As technologies grow more complex, uncertain, and powerful, our governance models must evolve accordingly. Nowhere is this clearer than in quantum computing, where uncertainty is a fundamental condition of the system itself. We must embrace uncertainty as a generative force. Applying this principle to quantum computing governance means creating frameworks that are inherently adaptive and responsive to new discoveries and changing contexts and could be transposed to other technologies.

This approach could be a particularly interesting initiative for the European Union to strengthen a joint industry policy based on the value framework of responsible technological development. Rather than attempting to categorize quantum applications into predefined risk levels, the EU should develop dynamic risk assessments that can evolve with the technology and create an alternating approach to libertarian and authoritarian models. By embracing the probabilistic und inherently uncertain nature of quantum phenomena in its governance structures, we can create an environment that encourages responsible development without imposing arbitrary limitations.

---

[1] Regulation (EU) 2024/1689 of the European Parliament and of the Council of 13 June 2024 laying down harmonised rules on artificial intelligence (AI Act)

[2] Named after the children's story "Goldilocks and the Three Bears" in which a girl tastes three different bowls of porridge to find out that she most prefers the one with just the right temperature. The concept has been used in different disciplines, e.g. in astronomy to describe the habitable zone around a star that is required for intelligent life.

[3] See, e.g. Regulation (EU) 2024/1689 laying down harmonized rules on artificial intelligence (EU AI Act). *Off. J. Eur. Union* (2024) defining the four different risk categories (prohibited; high; limited; minimal).

[4] C.f. the four Biosafety levels for pathogens in biomedical and clinical laboratories in United States Centers for Disease Control and Prevention (US CDC) & United States National Institutes of Health (US NIH), Biosafety in Microbiological and Biomedical Laboratories (BMBL), 6th edn. (2020). https://www.cdc.gov/labs/pdf/SF__19_308133-A_BMBL6_00-BOOK-WEB-final-3.pdf

[5] C.f. the five levels of driving automation in SAE International. *Taxonomy and Definitions for Terms Related to Driving Automation Systems for On-Road Motor Vehicles* (J3016_202104) (2021). https://www.sae.org/standards/content/j3016_202104/ ((accessed on June 9, 2025).

**Conflict of Interests:** The author(s) declared no potential conflicts of interest with respect to the research, authorship, and/or publication of this article'.

**Funding:** CHANSE, Collaboration of Humanities and Social Sciences in Europe Initiative (Q-SHIFT Project, https://www.quantumstateofworld.com). S.R.S. received funding from the European Research Council (ERC) under the projects NGBMI (759370) and BNCI2 (101088715), German Federal Ministry of Research, Technology and Space (BMFTR, 03ZU2110FC, 13N16486, 01UX2211) and the Einstein Stiftung Berlin

**Statements and Declarations:** Not applicable.